\documentclass[
twocolumn, 
 amsmath,amssymb,
 aps,
 prl,
longbibliography
]{revtex4}

\usepackage{graphicx}
\usepackage{dcolumn}
\usepackage{bm}

\begin{document}

\preprint{APS/123-QED}

\title{Metastable Hyperuniformity at Discontinuous Absorbing Transitions}

\author{Yusheng Lei}
 \affiliation{School of Chemistry, Chemical Engineering and Biotechnology, Nanyang Technological University, \\62 Nanyang Drive, 637459, Singapore}
\author{Ran Ni}\email{r.ni@ntu.edu.sg}
 \affiliation{School of Chemistry, Chemical Engineering and Biotechnology, Nanyang Technological University, \\62 Nanyang Drive, 637459, Singapore}

\date{\today}

\begin{abstract}
Nonequilibrium hyperuniformity can arise either as a steady-state property of driven active fluids or as a critical signature at continuous absorbing transition points in two and three dimensions. Whether analogous structural order exists near discontinuous absorbing transitions, and what mechanism generates it, remains unclear. Here, we show that discontinuous absorbing transitions generically host a metastable hyperuniform regime near the stability limit. Using a facilitated Manna model without center-of-mass conservation, we find anomalous scaling $S(k\to0)\sim k^{1.2}$, which appears only near the metastable regime and disappears both deep in the active phase and in the absorbing phase. This scaling is robust in both two and three dimensions, in contrast to critical hyperuniformity at continuous absorbing transitions. We further formulate a minimal conserved Reggeon field theory that reproduces the same metastable hyperuniform regime and anomalous scaling, demonstrating that the phenomenon does not rely on microscopic update rules but arises from the interplay of nonlinear activation, multiplicative demographic noise, and conserved diffusive fluctuations. These results identify metastable hyperuniformity as a generic pseudo-critical structural signature of discontinuous absorbing transitions coupled to a conserved density.\end{abstract}


\maketitle


Hyperuniformity, a structural feature introduced by Torquato and Stillinger~\cite{torquato2003local,torquato2016hyperuniformity,torquato2018hyperuniform}, characterizes a state of matter that anomalously suppresses the large-scale density fluctuation. Specifically, a system is hyperuniform if its structure factor $S(k)$ vanishes in the infinite-wavelength limit, i.e., the wave vector $k \to 0$. While perfect crystals and quasicrystals are canonical examples of ordered hyperuniformity, a number of  disordered hyperuniform structures have been discovered over the past two decades. These remarkable states of matter possess hidden long-range order despite their isotropic and amorphous nature, and have been identified in physical and biological systems ranging from jammed granular media~\cite{donev2005unexpected} to avian photoreceptor patterns~\cite{jiao2014avian}, leaf vein networks~\cite{liu2024universal}, and dryland vegetation patterns~\cite{hu2025causes}, etc. These disordered hyperuniform structures exhibit unique physical properties, such as isotropic photonic bandgaps~\cite{florescu2009designer,man2013isotropic}, drawing significant attention in materials science and related fields.

Recent research has been increasingly focused on generating dynamic hyperuniform structures out of equilibrium using both simulations and experiments~\cite{lei2025non,hexner2017noise,lei2019nonequilibrium,lei2019hydrodynamics,li2025fluidization,mitra2021hyperuniformity,zheng2024universal,wilken2023spatial,huang2021circular,oppenheimer2022hyperuniformity,zhang2022hyperuniform,weijs2015emergent,guo2026diffusion,hexner2015hyperuniformity,corte2008random,de2024hyperuniformity,anand2026emergent,backofen2024nonequilibrium,jack2015hyperuniformity,wang2025hyperuniform,wang2018hyperuniformity}, where hyperuniformity emerges either within the steady states of driven-dissipative systems~\cite{lei2019nonequilibrium,hexner2017noise,lei2019hydrodynamics,maire2025hyperuniformity,kuroda2023microscopic,li2025fluidization,Lei2023Spherical,anand2026emergent}, or at the critical point of systems undergoing continuous absorbing transitions~\cite{wiese2024hyperuniformity,hexner2015hyperuniformity,wilken2020hyperuniform,chen2024emergent,tjhung2015hyperuniform,wilken2021random,galliano2023two,ma2019hyperuniformity}.
While hyperuniformity at the critical points of continuous absorbing transitions in two and three dimensions has been well documented, its emergence near discontinuous absorbing transitions was only observed recently in systems with center-of-mass conservation~\cite{lei2021barrier,lei2023howHUfreeze,maire2025dynamical}. In these systems, hyperuniform structures with scaling $S(k\to 0)\sim k^{1.2}$ appear near the stability limit of non-equilibrium hyperuniform fluids that otherwise exhibit $S(k\to 0)\sim k^2$. However, the underlying physics of this new hyperuniform state remains unresolved: it is unclear whether it represents a structural crossover toward the absorbing state or a distinct emergent pseudo-critical phenomenon.
To this end, we perform extensive numerical simulations of a generic lattice model without center-of-mass conservation, namely a modified Manna model with a facilitation mechanism that renders the absorbing transition discontinuous. We find the robust emergence of a metastable active hyperuniform state near the stability limit, exhibiting structure factor scaling $S(k \to 0) \sim k^{1.2}$. 
Similar to critical hyperuniformity, which emerges near the critical point of a continuous absorbing transition, the metastable hyperuniform state appears only near the stability limit of the discontinuous absorbing transition; sufficiently far above or below this regime, the system always becomes non-hyperuniform. However, unlike critical hyperuniformity, whose scaling exponent depends on spatial dimensionality, we find that the metastable hyperuniform state exhibits the same scaling, $S(k\to 0)\sim k^{1.2}$, in both two and three dimensions.
To understand the underlying physics, we propose a minimal model with reaction rules, where activation is strongly favored in particle-rich regions inducing discontinuous absorbing transitions. By constructing and numerically simulating the corresponding Reggeon field theory, we show that the emergent metastable hyperuniformity is not tied to microscopic update rules, but arises generically in discontinuous absorbing transitions coupled to a conserved field.
Specifically, our theory shows that hyperuniformity in metastable states emerges from the combined effects of multiplicative demographic noise, nonlinear dynamics, and weak effective noise near the stability limit of a discontinuous absorbing transition, revealing a deep connection between hyperuniformity and the fundamental nature of discontinuous absorbing transitions.


\begin{figure*}[htb!]
    \centering
    \twocolumngrid
    \includegraphics[width=0.95\linewidth]{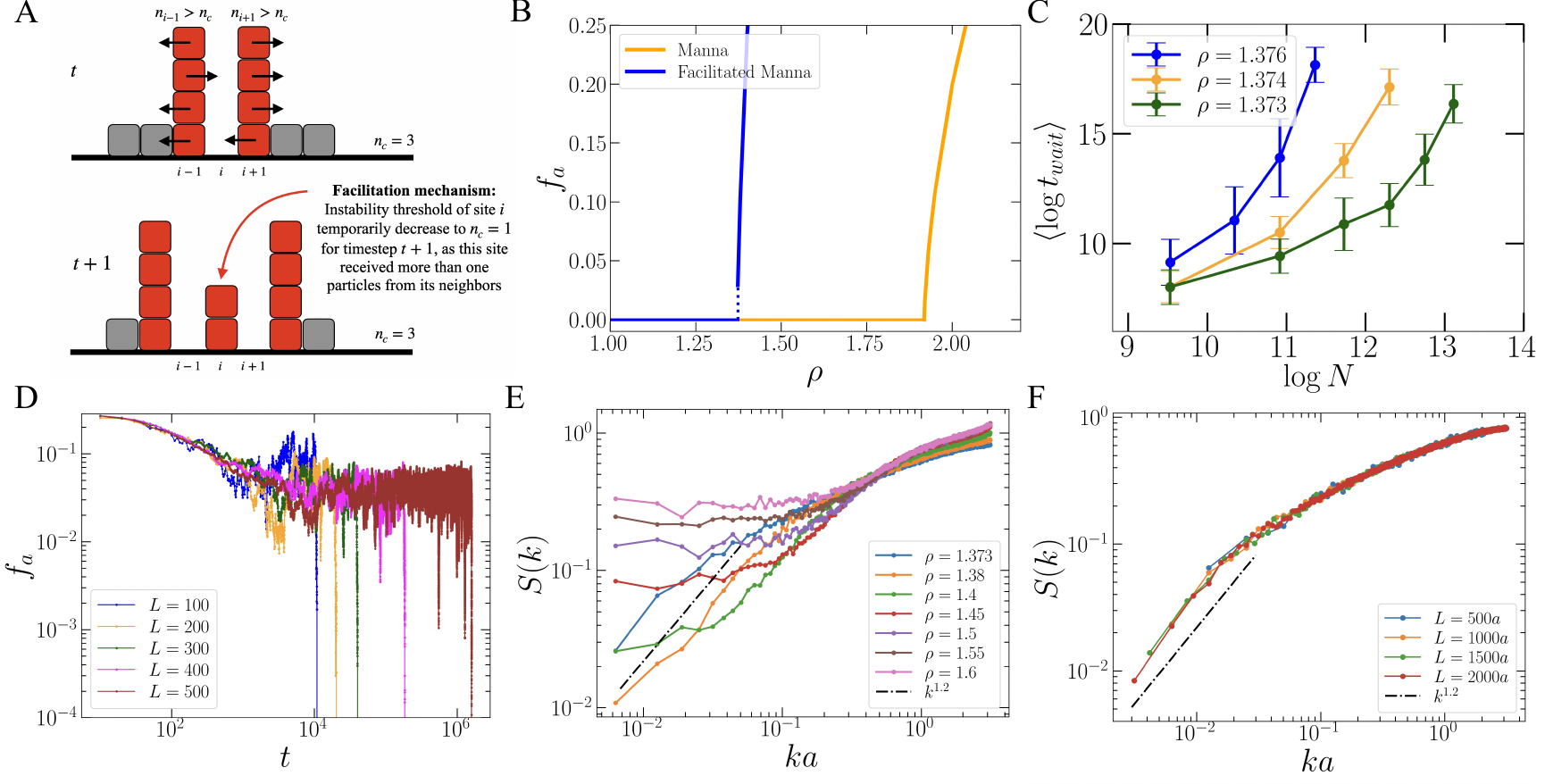}
    \caption{Facilitated Manna model in 2D. (A) Schematic diagram of the facilitated Manna model. Red and grey colors represent active and passive particles, respectively. (B) Steady state active particle fraction $f_a$ as a function of particle density $\rho$ for Manna model with and without Facilitation mechanism. (C) Surviving time of metastable state $\log(t_{wait})$ as a function of the system size $\log(N)$, where error bars are standard deviations. (D) The time evolution of the active particle fraction $f_a(t)$ in the metastable hyperuniform state of different sizes at $\rho=1.373$. (E) Structure factor of the metastable active states near the stability limit in Facilitated Manna model with system size $L=1000a$. (F) The structure factor $S(k)$ of the metastable hyperuniform state for different sizes at $\rho=1.373$. }
    \label{fig:1}
\end{figure*}


Similar to the original Manna model \cite{hinrichsen2000non,hexner2015hyperuniformity}, we consider a $d$-dimensional lattice where each site $i$ holds $n_i$ particles. We denote a site and the particles on it as active when $n_i> n_c$, with a certain threshold $n_c$, and in this work we choose $n_c=3$. 
At each timestep, only active sites are updated: all particles on active sites are randomly redistributed to neighboring sites. In addition to original Manna rules, as shown in Fig.~\ref{fig:1}(A), we adopt a facilitation mechanism that when a site $i$ receives more than one particle from its neighbors, the threshold $n_c$ of this site temporarily decreases to $n_c=1$ for that timestep, which effectively promotes activity in particle-rich regions, similar as in \cite{di2016self}. Notably, such activity induced redistribution is asymmetric, and active particles are redistributed randomly with the probability $p=1/2d$. This breaks the center-of-mass conservation while still conserving the total number of particles. Consequently, at low density $\rho$, the system falls into the absorbing phase with no active sites. At high density, it self-organizes to an active state where a finite density of active particles is maintained.

As shown in Fig.~\ref{fig:1}(B), compared to the original Manna model, the facilitation effect enables systems at lower densities to stay in an active state, and the order parameter, i.e., the fraction of particles on active sites $f_a \equiv \rho_A/\rho$, exhibits a discontinuous jump around $\rho \approx 1.373$, marking the stability limit of the metastable active state for a system size of $L=1000a$ with $a$ the lattice spacing. As shown in Fig.~\ref{fig:1}(C),  the observed metastable states exhibit highly anomalous system size dependence, and the waiting time of absorbing phase transition near the stability limit increases as the system size increases. We attribute this to a purely kinetic effect: while the system is metastable, it becomes kinetically stable at macroscopic scales, caused by large-scale fluctuation suppression, as observed in discontinuous absorbing transitions in similar models with center-of-mass conservation~\cite{lei2023howHUfreeze,maire2025dynamical}. With increasing the system size, the metastable active regime extends further toward the absorbing state, pushing the stability limit to lower values alongside a decreasing $f_a$. 
Nevertheless, we emphasize that the transition remains fundamentally discontinuous. As shown in Fig.~\ref{fig:1}(D), the time evolution of the metastable state near the stability limit lacks the standard finite-size scaling, which is typical in continuous absorbing transitions~\cite{lei2019hydrodynamics,hinrichsen2000non}, thereby confirming its discontinuous nature.

As shown in Fig.~\ref{fig:1}(E) and (F), the structure factor of the metastable active state exhibits a distinct hyperuniform scaling $S(k \to 0) \sim k^{1.2}$ near the stability limit. Due to the lack of a center-of-mass conservation, when the system moves deeper into the active state with larger $f_a$ and further away from the metastable region, this $k^{1.2}$ hyperuniform scaling gradually disappears, and a low-$k$ plateau arises in $S(k)$. As the same hyperuniform scaling of metastable states also appears in other systems undergoing discontinuous absorbing transitions ~\cite{lei2023howHUfreeze,maire2025dynamical}, this suggests that $k^{1.2}$ hyperuniform scaling of the metastable state is likely universal, rooted in the nature of the discontinuous transition. 
Furthermore, the emergent $k^{1.2}$ scaling appears to be distinct from the hyperuniformity associated with conventional active hyperuniform states: it disappears deep within the active phase and instead arises only near the metastable stability limit. This indicates that the scaling is tied to metastability itself, serving as a kinetic pseudo-critical signature of the discontinuous absorbing transition.


To understand the mechanism governing the metastable state and the discontinuous absorbing transition, we coarse-grain our microscopic lattice dynamics into effective local reaction rules~\cite{hinrichsen2000non,janssen2016directed,ma2025hyperuniformity} designed to capture the generic feature of systems undergoing discontinuous absorbing transitions. We propose minimal reactions that represent the  facilitation effect, where activation ($P \to A$) is strongly favored in $A$-rich regions. The local reactions for active ($A$) and passive ($P$) particles are 
\begin{equation*}
        A \xrightarrow[]{\mu} P, 
        A+P \xrightarrow[]{\kappa} 2A, 
        2A+P  \xrightarrow[]{\lambda} 3A,
\end{equation*}
where $\mu$ is the deactivation rate. The three-body activation rate $\lambda$ dominates over the two-body activation rate $\kappa$, leading to  strong cooperative activation in the system. In this system, only active particles can diffuse, and total particle number of $A$ and $P$ is conserved.

Similar to phenomenological descriptions proposed for systems exhibiting self-organized bistability~\cite{di2016self}, the facilitation effect naturally yields a third-order nonlinearity. This leads to a region of metastability in the mean-field dynamics, which drives the nature of the absorbing transition to be discontinuous. Denoting the active and passive particle densities as $\rho_A$ and $\rho_P$, respectively, and utilizing the conserved total density $\rho = \rho_A + \rho_P$, the mean-field equation for $\rho_A$ is given by
\begin{equation}
    \begin{aligned}
        \dot{\rho}_A&=-\mu\rho_A+\kappa\rho_A\rho_P+\lambda\rho_A^2\rho_P \\
        &=(\kappa\rho-\mu)\rho_A-(\kappa-\lambda\rho)\rho_A^2-\lambda\rho_A^3,\\
    \end{aligned}
\end{equation}
where the reaction rates $\kappa,\mu,\lambda>0$, and $\kappa-\lambda\rho<0$ ensures the existence of a discontinuous transition. Note that while the actual facilitated Manna model may involve higher-order nonlinearities, our minimal third-order terms are sufficient to govern the metastability and capture essential dynamical and structural properties of such discontinuous absorbing transitions.

To extend this deterministic mean-field description into a full stochastic field theory capable of capturing spatial fluctuations, we employ the Doi-Peliti coherent-state path integral formalism~\cite{hinrichsen2000non,janssen2016directed,ma2025hyperuniformity,wiese2016coherent}. We denote a microscopic configuration of the system as $|n\rangle \equiv |\{n_{A,i}\}, \{n_{P,i}\}\rangle$, where $n_{A,i}$ and $n_{P,i}$ are the amounts of particles on site $i$. Considering  $\left | \psi(t) \right \rangle=\sum_n P(n,t)|n\rangle$ with probability density $P(n,t)$ of state $|n\rangle$, the master equation of the system can be written as 
\begin{equation} 
    \partial_t\left | \psi(t) \right \rangle=-{\mathcal H} \left | \psi(t) \right \rangle
\end{equation}
with the Hamiltonian operator
\begin{equation}
\label{hamiltonian}
\begin{aligned}
    {\mathcal  H}=&D\sum_{(i,j)}(\hat a_i-\hat a_j\big)( a_i- a_j)-\sum_i\left[\mu(\hat p_i-\hat a_i)a_i\right.\\
    &\left.+\kappa\left(\hat a_i^{2}-\hat a_i\hat p_i\right)a_ip_i+\lambda\left(\hat a_i^{3}-\hat a_i^{2}\hat p_i\right ) a_i^{2}p_i\right]
\end{aligned}
\end{equation}
where $D$ is the intrinsic diffusion constant of particle $A$, and the bosonic creation/annihilation operators $\left\{\hat  a_i, a_i \right\}$ for $A$, satisfying $ a_i|0\rangle=0,\, \hat a_i|n_{A,i}\rangle=|n_{A,i}+1\rangle,\,  a_i|n_{A,i}\rangle=n_{A,i}|n_{A,i}-1\rangle$. $\left\{ \hat p_i,p_i \right\}$ for $P$ at site $i$, satisfies $ p_i|0\rangle=0,\, \hat p_i|n_{P,i}\rangle=|n_{P,i}+1\rangle,\,  p_i|n_{P,i}\rangle=n_{P,i}|n_{P,i}-1\rangle$.
Taking the continuum limit and considering the corresponding eigenvalue fields of operators $\hat a$,$a$,$\hat p$,$p$, we apply the quasicanonical Grassberger (also called Cole-Hopf) transformation $\hat{a}=\exp \left(\tilde{n}_A\right)$, $a=n_A \exp \left(-\tilde{n}_A\right)$, $\hat{p}=\exp \left(\tilde{n}_P\right)$, $p=n_P \exp \left(-\tilde{n}_P\right)$~\cite{janssen2016directed,lefevre2007dynamics}, so that $n_A(\boldsymbol{x},t)=\hat aa$ and $n_P(\boldsymbol{x},t)=\hat pp$ are active and passive particle density fields. To capture the conservation of total number of particles, we perform linear variable substitutions to decouple the conserved and active fluctuating modes~\cite{janssen2016directed}: $n_A+n_P\rightarrow\rho$, $n_A\rightarrow\rho_A$, and response fields $\tilde n_A-\tilde n_P\rightarrow \tilde\rho_A$, $\tilde n_P\rightarrow \tilde \rho$. Truncating the full action $\mathcal{S}$ to quadratic order of the response fields $\{\tilde \rho_A,\tilde \rho\}$ to capture all necessary fluctuations near the phase transition, the reduced action functional is
\begin{equation}
    \begin{aligned}
        \mathcal S=&\int d^d x \int_{-\infty}^{+\infty} d t \left\{\tilde{\rho}_A  \partial_t \rho_A-\tilde{\rho}_A D\nabla^2 \rho_A+\tilde{\rho}  \partial_t \rho\right. \\ 
        & \left.-\tilde{\rho} D\nabla^2 \rho_A-\tilde{\rho}_A\left( (\kappa\rho-\mu)\rho_A-(\kappa-\lambda\rho)\rho_A^2-\lambda\rho_A^3\right) \right. \\ 
        &\left.+D\left( \nabla \left(\tilde\rho +\tilde\rho_A\right) \right)^2\rho_A -\frac{1}{2} \mu\left(\tilde{\rho}_A\right)^2 \rho_A\right\}
    \end{aligned} 
\end{equation}
Following the response-field path integral formalism, we derive the noise term of our field equation by transforming the quadratic response field terms $D\left( \nabla \left(\tilde\rho +\tilde\rho_A\right) \right)^2\rho_A$ and $\frac{1}{2} \mu\left(\tilde{\rho}_A\right)^2 \rho_A$, and the generating functional $\mathcal{Z}=e^{-\mathcal{S}}$ (see End Matter)
\begin{equation}
\label{generating_function}
    \begin{aligned}
        \mathcal{Z} =& \left\langle\exp  {\int d^d x \int_{-\infty}^{+\infty} d t  \left\{-\tilde{\rho}_A \partial_t \rho_A+\tilde{\rho}_A D\nabla^2 \rho_A-\tilde{\rho}  \partial_t \rho\right.}\right. \\
        &+\tilde{\rho} D\nabla^2 \rho_A+\tilde{\rho}_A\left( (\kappa\rho-\mu)\rho_A-(\kappa-\lambda\rho)\rho_A^2-\lambda\rho_A^3\right)\\
        &\left.\left. +\tilde{\rho}_A \sqrt{\mu \rho_A} \eta+\left(\tilde\rho +\tilde\rho_A\right)  \nabla\cdot \sqrt{ 2D\rho_A} \boldsymbol\Xi\right\}\right\rangle_{\eta,\,\boldsymbol\Xi}
    \end{aligned}
\end{equation}
where $\eta(\boldsymbol{x},t)$ and $\boldsymbol{\Xi}(\boldsymbol{x},t)$ are scalar and $d$-dimensional unit Gaussian white noises, respectively. $\langle \mathcal{O}[\eta,\boldsymbol \Xi] \rangle_{\eta,\boldsymbol\Xi} \equiv \int \mathcal{D}\eta\mathcal{D}\boldsymbol\Xi \, \mathcal{O}[\eta,\boldsymbol\Xi] \, { \exp\left( -\frac{1}{2} \int d^dx dt \, (\eta^2+\boldsymbol\Xi^2) \right) }$ with path integral measure $\mathcal{D}\eta$ absorbs the proper normalization such that $\langle 1 \rangle_{\eta,\boldsymbol\Xi} = 1$. Thus, the resulting Reggeon field equations are: 
\begin{equation}
\label{RFT}
    \begin{aligned}
        \partial_t \rho(\boldsymbol{x},t) =& D \nabla^2 \rho_A+\nabla\cdot \sqrt{2D\rho_A}\boldsymbol\Xi\\
        \partial_t \rho_A(\boldsymbol{x},t) =& (\kappa\rho-\mu)\rho_A-(\kappa-\lambda\rho)\rho_A^2-\lambda\rho_A^3+ D \nabla^2 \rho_A \\
        & + \sqrt{\mu\rho_A} \eta+\nabla\cdot \sqrt{2D\rho_A}\boldsymbol\Xi
    \end{aligned}
\end{equation}
To investigate the structural feature of the discontinuous absorbing  transition, we numerically integrate the field equations on a two-dimensional square lattice of $L=100a$ with periodic boundary conditions (see SI). We employ a random initial condition $\rho(\boldsymbol{x},0)=\rho_A(\boldsymbol{x},0)=\bar\rho(1+0.1\delta(\boldsymbol{x}))$, where $\delta(\boldsymbol{x})$ is drawn from a unit Gaussian distribution. The integration is performed with a diffusivity $D=0.01$ and a time step $\Delta t=0.05$. The reaction parameters are fixed at $\mu=1.4$, $\kappa=0.1$, and $\lambda=1.0$, chosen such that $\kappa-\lambda\bar\rho < 0$ to strictly ensure the emergence of discontinuous phase transition  in the system. For simplicity, we neglect the diffusive noise term in the evolution equation for $\rho_A$. Because as a higher-order spatial derivative, it scales as $\mathcal{O}(k^3)$ in Fourier space and is strictly subleading compared to the demographic noise $\eta$, thus not altering the macroscopic hyperuniform properties of the field $\rho(\boldsymbol x,t)$ (see End Matter).

\begin{figure}[htb!]
    \centering
    \includegraphics[width=0.9\linewidth]{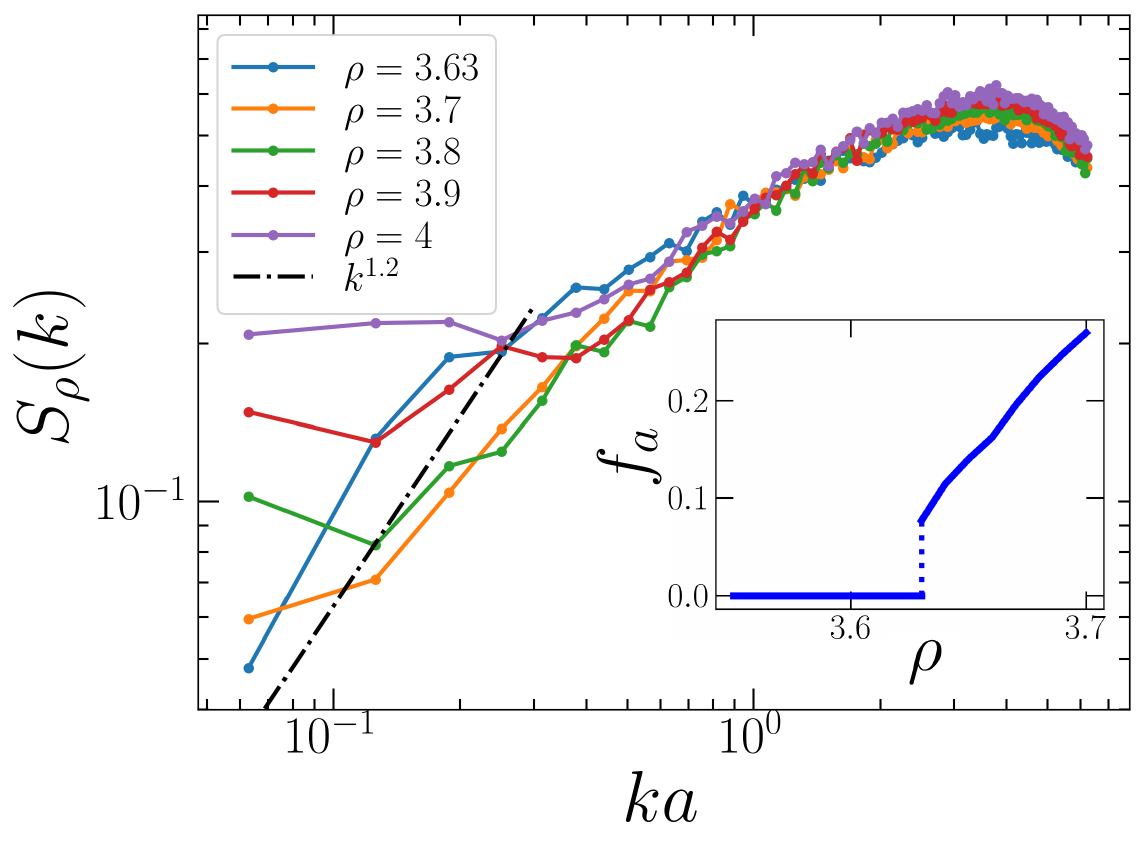}
    \caption{Reggeon field theory simulation in 2D with system size $L=100$, $D=0.01$, $\mu=1.4$, $\kappa=0.1$, $\lambda=1$. Structure factor near metastable active states. Inset: Steady state active particle fraction $f_a$ as a function of particle density $\rho$. }
    \label{fig:2}
\end{figure}

As shown in the inset of Fig.~\ref{fig:2}, we observe a discontinuous absorbing transition, where the active fraction $f_a=\rho_a/\rho$ jumps near $\rho\approx 3.63$. We measure the structure factor $S_\rho(k)$ of the density field, which also shows the hyperuniform scaling $S_{\rho}(k\to 0) \sim k^{1.2}$ near the stability limit of metastable active phase. With increasing the active fraction $f_a$, a plateau in $S_\rho(k)$ appears at small $k$, breaking the hyperuniform scaling as the system moves deeper into the active phase, which quantitatively agrees with the simulation of facilitated Manna model (Fig.~\ref{fig:1}). Therefore, our field theory demonstrates that the anomalous hyperuniform scaling $S_{\rho}(k) \sim k^{1.2}$ is not merely a product of specific lattice update rules, but rather a robust universal feature of discontinuous absorbing phase transitions driven by facilitation/barrier effects ~\cite{barrier_SI}. This universality is fundamentally governed by the interplay between the deterministic third-order nonlinearity, which drives the discontinuous transition, and the competing stochastic noises.

Crucially, our field-theoretic framework formalizes the physical mechanism regarding the loss of center-of-mass conservation. Near the stability limit of the metastable state, the active particle density $\rho_A$ is marginally small. Therefore, the deterministic third-order nonlinear term, the  demographic noise ($\sqrt{\rho_A}\eta$) and diffusive noise ($\nabla\cdot\sqrt{\rho_A}\boldsymbol{\Xi}$) collectively dominate the spatial structural fluctuation. However, when the system transitions deep into the active phase, $\rho_A$ increases substantially. The diffusive noise $\boldsymbol\Xi$, acting as an effective source of local large density fluctuations, gradually takes over. This diffusive noise effectively breaks the strict center-of-mass conservation constraint at the mesoscopic level, acting as a structural decorrelator that washes out the long-range hyperuniform order and leads to the observed low-$k$ plateau~\cite{de2024hyperuniformity,ma2025hyperuniformity}. Therefore, the hyperuniform $k^{1.2}$ scaling within a narrow metastable window of a discontinuous phase transition can be understood as an emergent behavior that balances the deterministic third-order nonlinearity, demographic noise, and the dominant conserved diffusive flux.

\begin{figure*}[htb!]
    \centering
    \twocolumngrid
    \includegraphics[width=0.75\linewidth]{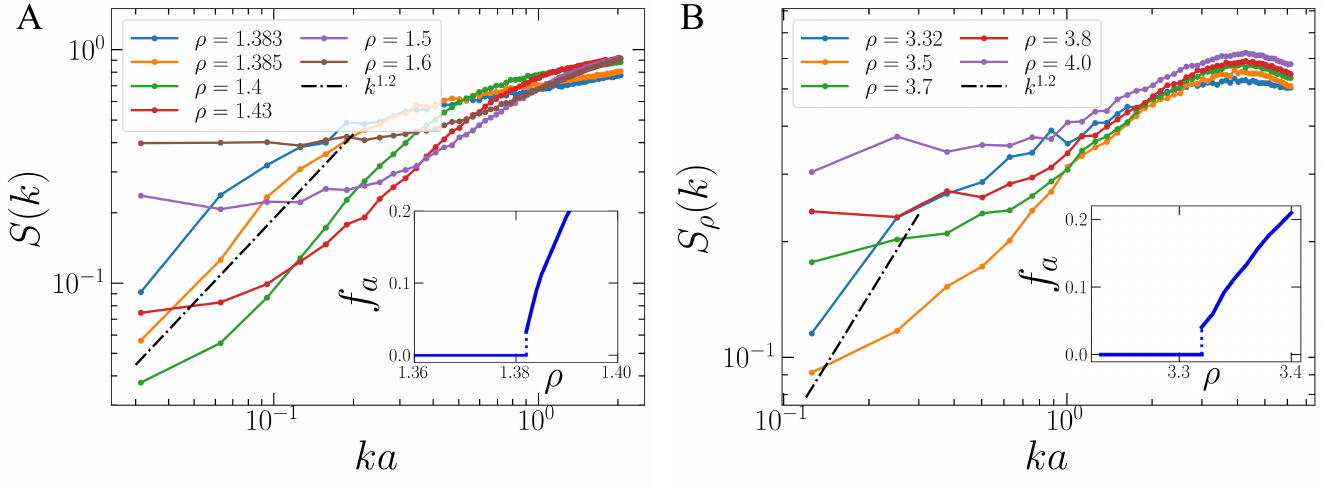}
    \caption{Comparison between Facilitated Manna model and the Reggeon field theory simulation in 3D. (A) Facilitated Manna model in 3D with system size $L=200$, $n_c=3$: Structure factor near metastable active states. Inset: Steady state active particle fraction $f_a$ as a function of particle density $\rho$; (B) Reggeon field theory simulation in 3D with system size $L=50$, $D=0.01$, $\mu=1.4$, $\kappa=0.1$, $\lambda=1$: Structure factor near metastable active states. Inset: Steady state active particle fraction $f_a$ as a function of particle density $\rho$.}
    \label{fig:3}
\end{figure*} 


Furthermore, to investigate the role of spatial dimensionality in this anomalous hyperuniformity, we performed simulations of both the facilitated Manna model and the corresponding Reggeon field theory in three dimensions (3D). As shown in Fig.~\ref{fig:3}, decreasing the system density yields a clear discontinuous absorbing phase transition in $f_a$, similar to the 2D case. By measuring the structure factor near the stability limit, we observe that the metastable active state exhibits the exact same hyperuniform scaling $S_\rho(k) \sim k^{1.2}$. Similarly, a low-$k$ plateau emerges as the system transitions deeper into the active phase. Crucially, unlike the critical hyperuniformity associated with continuous absorbing phase transitions, which typically exhibits a strong dependence on the spatial dimension, the anomalous metastable hyperuniformity driven by this discontinuous transition appears to be robust and independent of dimensionality.

In summary, we have identified a metastable hyperuniform state that emerges near the stability limit of discontinuous absorbing phase transitions. Using a facilitated Manna model without center-of-mass conservation, we show that this state exhibits $S(k\to 0)\sim k^{1.2}$, which appears only within a narrow metastable regime, and remains robust in both two and three dimensions. We further formulate a minimal conserved Reggeon field theory that reproduces this behavior, showing that metastable hyperuniformity does not rely on microscopic update rules but arises from the generic interplay of nonlinear activation, multiplicative demographic noise, and conserved diffusive fluctuations. These results establish metastable hyperuniformity as a generic pseudo-critical structural signature of discontinuous absorbing phase transitions coupled to a conserved density.

\begin{acknowledgments}
This work was financially supported by the Academic Research Fund from the Singapore Ministry of Education (RG151/23, and RG88/25) and the National Research Foundation, Singapore, under its 29th Competitive Research Program (CRP) Call (NRF-CRP29-2022-0002).
\end{acknowledgments}

\paragraph{Data availability}
The data that support the findings of
this article are not publicly available. The data are available
from the authors upon reasonable request.

\bibliography{paper}

\begin{thebibliography}{50}
\bibitem[{bar()}]{barrier_SI}
\bibinfo{note}{Discussions about the barrier effect can be found in SI, and we
  show that the reaction rule respresenting barrier effect should share the
  similar form of Reggeon field equations with third order $\rho_A$
  nonlinearity.}
\end{thebibliography}

\appendix
\section{End Matter}
\subsection{Derivation of the Reggeon field theory}

From Eq.~\ref{hamiltonian}, the corresponding action can be formally given by $\mathcal{S}[\{\hat a,a,\hat p, p\}] = \int d^d x \int_{-\infty}^{+\infty} d t \left[ (\hat{a}-1)\partial_t a + (\hat{p}-1)\partial_t p + \mathcal{H}(\hat{a}, a, \hat{p}, p) \right]$, where the temporal derivative terms indicate the probability-conserving evolution, and the Hamiltonian $\mathcal{H}$ captures all local reaction and diffusion dynamics. Thus, we have
\begin{equation}
    \begin{aligned}
        \mathcal S =&\int d^d x \int_{-\infty}^{+\infty} d t\left\{(\hat{a}-1) \partial_t a+(\hat{p}-1) \partial_t p+D \nabla \hat{a} \cdot \nabla a\right.\\
        & \left.-\mu(\hat{p}-\hat{a}) a-\kappa(\hat{a}^2-\hat{a} \hat{p}) a p- \lambda(\hat a^{3}-\hat a^{2}\hat p)a^{2}p\right\}
    \end{aligned}
\end{equation}
where the coherent fields $\hat a$,$a$,$\hat p$,$p$ correspond to the coherent-state eigenvalues fields of the bosonic creation/annihilation operators of the Fock space, with initial and final conditions $\hat a(\boldsymbol x,\infty)=\hat p (\boldsymbol x,\infty)= 1$ and $a(\boldsymbol x,-\infty)=p(\boldsymbol x,-\infty)= 0$~\cite{janssen2016directed}. It is worth noting that, compared to the standard conserved directed percolation (CDP) model described in \cite{janssen2016directed,hinrichsen2000non}, the fundamental distinction here is the additional term $\lambda(\hat a^{3}-\hat a^{2}\hat p)a^{2}p$  originating from the facilitation reaction $2A+P \xrightarrow{\lambda} 3A$.
 
In order to transform the action to physical particle density fields, we apply a quasicanonical Grassberger (also called Cole-Hopf) transformation $\hat{a}=\exp \left(\tilde{n}_A\right)$, $a=n_A \exp \left(-\tilde{n}_A\right)$, $\hat{p}=\exp \left(\tilde{n}_P\right)$, $p=n_P \exp \left(-\tilde{n}_P\right)$~\cite{janssen2016directed,lefevre2007dynamics}, so that $n_A(\boldsymbol{x},t)=\hat aa$ and $n_P(\boldsymbol{x},t)=\hat pp$ are active and passive particle density fields. Then, the action is
\begin{equation}
    \begin{aligned}
        \mathcal S=&\int d^d x d t\left\{\tilde{n}_A \partial_t n_A+\tilde{n}_P \partial_t n_P+D\left[\nabla \tilde{n}_A \cdot \nabla n_A\right. \right. \\ 
        &\left.\left.-n_A\left(\nabla \tilde{n}_A\right)^2\right]-\left[e^{\tilde{n}_A-\tilde{n}_P}-1\right]\left(\kappa n_A n_P+\lambda n_P n_A^2\right)\right.\\
        &\left.-\left[e^{\tilde{n}_P-\tilde{n}_A}-1\right] \mu n_A\right\}
    \end{aligned}
\end{equation}

Next, to explicitly capture the global conservation of the total particle number and evolution of the active particle density in our field theory, we perform linear variable substitutions to decouple the conserved and active fluctuating modes~\cite{janssen2016directed}: $n_A+n_P\rightarrow\rho$, $n_A\rightarrow\rho_A$, and response fields $\tilde n_A-\tilde n_P\rightarrow \tilde\rho_A$, $\tilde n_P\rightarrow \tilde \rho$, so that $\rho(\boldsymbol x,t),\rho_A(\boldsymbol x,t)$ represent the total and active particle density fields, respectively. Truncating the full action $\mathcal{S}$ to quadratic order of the response fields $\{\tilde \rho_A,\tilde \rho\}$ to capture all necessary fluctuations near the phase transition, the reduced action functional would be
\begin{equation}
    \begin{aligned}
        \mathcal S=&\int d^d x \int_{-\infty}^{+\infty} d t \left\{\tilde{\rho}_A  \partial_t \rho_A-\tilde{\rho}_A D\nabla^2 \rho_A+\tilde{\rho}  \partial_t \rho\right. \\ 
        & \left.-\tilde{\rho} D\nabla^2 \rho_A-\tilde{\rho}_A\left( (\kappa\rho-\mu)\rho_A-(\kappa-\lambda\rho)\rho_A^2-\lambda\rho_A^3\right) \right. \\ 
        &\left.+D\left( \nabla \left(\tilde\rho +\tilde\rho_A\right) \right)^2\rho_A -\frac{1}{2} \mu\left(\tilde{\rho}_A\right)^2 \rho_A\right\}
    \end{aligned} 
\end{equation}
Then by taking the generating functional Eq.~\ref{generating_function}
, we can obtain Eq.~\ref{RFT} in the main text by
\begin{equation}
    \begin{aligned}
        \left.\frac{\delta \mathcal Z\left [ \left \{\rho_A,\tilde \rho_A, \rho, \tilde \rho\right\}\right ]}{\delta \tilde{\rho}(\boldsymbol x,t)}\right|_{\tilde{\rho}=0} & = 0\\
        \left.\frac{\delta \mathcal Z\left [ \left \{\rho_A,\tilde \rho_A, \rho, \tilde \rho\right\}\right ]}{\delta \tilde{\rho}_A(\boldsymbol x,t)}\right|_{\tilde{\rho}_A=0} & = 0
    \end{aligned}
\end{equation}

\section{The effect of diffusive noise in field equation of $\rho_A$}

In our Reggeon field theory simulations, we neglect the diffusive noise in the evolution equation for $\rho_A$. In this section, we demonstrate through linearization and adiabatic substitution into the equation for $\rho$ that this noise acts as a higher-order spatial derivative term compared to the dominant demographic and conserved noises. Consequently, it is negligible at large scales and does not alter the structural properties of interest, i.e., the hyperuniformity of the total density field $\rho(\boldsymbol{x},t)$.

In steady-state approximation, we can take the linearization with $\rho_A=\bar\rho_A+\delta\rho_A$ and $\rho =\bar\rho+\delta\rho$. Then, adopting Taylor expansions and neglecting higher order nonlinear terms, the deterministic terms $F(\rho_A,\rho)\equiv(\kappa\rho-\mu)\rho_A-(\kappa-\lambda\rho)\rho_A^2-\lambda\rho_A^3$ would be
\begin{equation}
    \begin{aligned}  F(\rho_A,\rho)=&F_0 +\alpha_0\delta \rho_A+\beta_0\delta \rho+\mathcal{O}\left(\delta \rho^2\right)+\mathcal{O}\left(\delta \rho_A^2\right)
    \end{aligned}
\end{equation}
where $\alpha_0\equiv\left(\partial_{\rho_A} F\right)|_{\bar \rho_A,\bar \rho}=(\kappa\bar\rho-\mu)-2(\kappa-\lambda\bar\rho)\bar\rho_A-3\lambda\bar\rho_A^2$, $\beta_0\equiv\left(\partial_{\rho} F\right)|_{\bar \rho_A,\bar \rho}=\kappa \bar\rho_A+\lambda\bar\rho_A^2$, and $F_0\equiv F(\bar\rho_A,\bar\rho)$.

Expanding the noise terms to leading order yields $\sqrt{\mu\rho_A}\eta \simeq \sqrt{\mu\bar{\rho}_A}\eta$ and $\nabla \cdot (\sqrt{2D\rho_A} \boldsymbol{\Xi}) \simeq \sqrt{2D\bar{\rho}_A} \nabla \cdot \boldsymbol{\Xi}$. Thus, the linearized coupled equations are
\begin{equation}
    \begin{aligned} 
        \partial_t \delta\rho =& D\nabla^2\delta\rho_A + \sqrt{2D\bar\rho_A}\nabla\cdot\boldsymbol{\Xi}, \\ 
        \partial_t \delta\rho_A =& F_0 + \alpha_0\delta\rho_A + \beta_0\delta\rho + D\nabla^2\delta\rho_A + \sqrt{\mu\bar\rho_A}\eta \\
        &+ \sqrt{2D\bar\rho_A}\nabla\cdot\boldsymbol{\Xi}'.
\end{aligned}
\end{equation}

Because the active density $\rho_A$ is not a conserved quantity, it acts as a fast variable that rapidly relaxes to a local steady state slaved to the slow conserved field $\rho$. We apply the adiabatic approximation by setting $\partial_t \delta\rho_A = 0$. Transforming to Fourier $(\boldsymbol{k},\omega)$-space, we obtain
\begin{equation}
    \delta \rho_A(\boldsymbol{k},\omega) = \frac{-F_0 - \beta_0 \delta \rho - \sqrt{\mu\bar{\rho}_A} \eta - \sqrt{2D\bar{\rho}_A} (i \boldsymbol{k} \cdot \boldsymbol{\Xi}')}{\alpha_0 - D k^2}
\end{equation}
Inserting this slaved solution into the Fourier-transformed equation for the slow variable $-i \omega \delta \rho = -D k^2 \delta \rho_A + \sqrt{2D\bar{\rho}_A} (i \boldsymbol{k} \cdot \boldsymbol{\Xi})$, we get
\begin{equation}
\begin{aligned}
    -i \omega \delta \rho =& \frac{D k^2 F_0}{\alpha_0 - D k^2} + \frac{D k^2 \beta_0}{\alpha_0 - D k^2} \delta \rho + \frac{D l^2 \sqrt{\mu \bar{\rho}_A}}{\alpha_0 - D k^2} \eta \\
    & + \sqrt{2D\bar{\rho}_A} (i\boldsymbol{k} \cdot \boldsymbol{\Xi})+ \frac{D k^2 \sqrt{2D \bar{\rho}_A}}{\alpha_0 - D k^2} (i \boldsymbol{k} \cdot \boldsymbol{\Xi}')
\end{aligned}
\end{equation}

Taking long-wavelength limit, i.e., $k\rightarrow 0$,  we can reverse the equation back to real space, the effective macroscopic evolution of the density field $\rho$ obeys 
\begin{equation}
    \begin{aligned}
        \partial_t \rho&\simeq D_{\mathrm {eff }} \nabla^2 \rho+\sigma_\eta \nabla^2 \eta+\sigma_{\Xi, 1} \nabla \cdot \boldsymbol{\Xi}+\sigma_{\Xi, 2} \nabla^2 \nabla \cdot \boldsymbol{\Xi}'\\
        &\simeq D_{\mathrm {eff }} \nabla^2 \rho+\sigma_\eta \nabla^2 \eta+\sigma_{\Xi, 1} \nabla \cdot \boldsymbol{\Xi}+\mathcal O\left(\nabla^3\right)
    \end{aligned}
\end{equation}
where we have expanded the prefactor for small $k$ as $\frac{D k^2 \beta_0}{\alpha_0 - D k^2} \simeq D_{\mathrm{eff}} k^2 + \mathcal{O}(k^4)$ with the effective diffusion constant $D_{\mathrm{eff}} = -D \beta_0 / \alpha_0$. Similarly, the noise coefficients are identified as $\sigma_\eta = -D \sqrt{\mu \bar{\rho}_A} / \alpha_0$, $\sigma_{\Xi, 1} = \sqrt{D \bar{\rho}_A}$, and $\sigma_{\Xi,2} = -D \sqrt{D \bar{\rho}_A} / \alpha_0$.

Here, $\sigma_\eta \propto \sqrt{\mu\bar{\rho}_A}$ originates from the demographic noise, while $\sigma_{\Xi, 2} \propto \sqrt{2D\bar{\rho}_A}$ originates from the diffusive noise in the $\rho_A$ equation. Crucially, the direct diffusive flux in the $\rho$ equation contributes the leading-order conserved noise $\sigma_{\Xi, 1} \nabla \cdot \boldsymbol{\Xi}$, which scales as $\mathcal{O}(\nabla)$. The demographic noise contributes an effective Laplacian noise scaling as $\mathcal{O}(\nabla^2)$. In contrast, the diffusive noise mapped from the $\rho_A$ equation becomes a highly subleading term $\nabla^2 (\nabla \cdot \boldsymbol{\Xi}')$ scaling as $\mathcal{O}(\nabla^3)$. Therefore, at macroscopic scales ($k \to 0$), the contribution from $\mathcal{O}(\nabla^3)$ term to the structure factor and dynamic fluctuations is negligible.

\end{document}